\documentclass[a4paper,12pt]{article}

\usepackage{jheppub} 
                     
\usepackage{amsmath}
  \usepackage{amssymb}
  \usepackage{amsfonts}
\usepackage{dsfont}
\usepackage[colorlinks=true]{hyperref}
\usepackage[capitalize]{cleveref}
  \Crefname{section}{Sec.}{Secs.}
  \Crefname{appendix}{App.}{Apps.}
  \Crefname{table}{Tab.}{Tabs.}
\usepackage{physics}
\usepackage{mathtools}
\usepackage{graphicx}
\usepackage{slashed}
\usepackage{epstopdf}
\usepackage{tensor}
\usepackage{comment}
\usepackage{caption}
\usepackage{subcaption}
\usepackage[shortlabels,inline]{enumitem}  
\usepackage{siunitx}

\title{\boldmath Anomaly Detection in the Presence of Irrelevant Features}
\author[a,1]{Marat Freytsis\note{Present affiliation: Anthropic, San Francisco, California 94960, USA},}
\author[a]{Maxim Perelstein,}
\author[a]{Yik Chuen San}

\affiliation[a]{Department of Physics, LEPP, Cornell University, Ithaca, NY 14853, USA}

\abstract{Experiments at particle colliders are the primary source of insight into physics at microscopic scales. Searches at these facilities often rely on optimization of analyses targeting specific models of new physics. Increasingly, however, data-driven model-agnostic approaches based on machine learning are also being explored. A major challenge is that such methods can be highly sensitive to the presence of many irrelevant features in the data. This paper presents Boosted Decision Tree (BDT)-based techniques to improve anomaly detection in the presence of many irrelevant features. First, a BDT classifier is shown to be more robust than neural networks for the Classification Without Labels approach to finding resonant excesses assuming independence of resonant and non-resonant observables. Next, a tree-based probability density estimator using copula transformations demonstrates significant stability and improved performance over normalizing flows as irrelevant features are added. The results make a compelling case for further development of tree-based algorithms for more robust resonant anomaly detection in high energy physics.}

\begin{document}
\maketitle
\flushbottom

\section{Introduction}
\label{sec:intro}

Experiments at high-energy colliders, such as the Large Hadron Collider (LHC), continue to be the primary source of information about the nature of physics at the microscopic scales.
A major task of the current and future experiments is to search for deviations from the Standard Model (SM) of particle physics.
Traditionally, such searches are performed by assuming a particular model for physics beyond the Standard Model (BSM), and optimizing the event selection and statistical analysis to obtain maximum sensitivity to the new physics signal in the presence of the SM background.
Increasingly, these are supplemented with data-driven methods which minimize model-dependent assumptions about the structure of deviations from the SM, with machine-learning (ML) based approaches the primary driver of such searches~\cite{Kasieczka:2021xcg,Aarrestad:2021oeb}.

The enormous size and complexity of the data sets collected by collider experiments currently preclude conducting a search for ``anything that doesn't look like the SM'' in the full data set at once.
Even if it were possible in principle, the dependence of collider analyses on complex simulations to interpret measured signals would make such an approach extremely sensitive to mismodelling errors at all stages of the simulation chain.
Instead, recent work focuses on a simpler task of anomaly detection when localized with respect to a particular variable~\cite{Collins:2018epr,Heimel:2018mkt,Farina:2018fyg,Collins:2019jip,Nachman:2020lpy,Andreassen:2020nkr,Benkendorfer:2020gek,Hallin:2021wme,Raine:2022hht,Hallin:2022eoq,Golling:2022nkl}.
A well-studied benchmark example, starting with the work of~\cite{Collins:2018epr,Collins:2019jip}, is a search for a dijet resonance, in which the signal jets are produced by a boosted resonance decay which is imprinted in non-trivial jet substructure.
ML techniques allow for searches of anomalous events with such topology, without making strong model-dependent assumptions about the new physics model that gives rise to this signal.
The original algorithm used the Classification Without Labels (CWoLa) approach~\cite{Metodiev:2017vrx} to train a neural-network (NN) classifier on between a signal region and the background-dominated data sample from the regions in the invariant mass $m_{JJ}$ outside of the signal window, and applied it to search for anomalous events in the signal region.
This approach requires that the features distinguishing signal and background be uncorrelated with $m_{JJ}$, which is not always the case in real-world applications.
To circumvent this problem, algorithms such as ANODE~\cite{Nachman:2020lpy} and CATHODE~\cite{Hallin:2021wme} were developed to detect anomalies based on probability density estimation.

A serious issue that can hinder practical applications of the automated anomaly detection methods is the rapid deterioration of performance with growing dimensionality of data space.
Typically, collider data contains some observables (or \emph{features}) that are relevant for discriminating signal and background, and a number of observables whose distribution is very similar in the signal and background samples.
In a true model-agnostic search, one rarely has the privilege of knowing what features are important beforehand, and inevitably many of the included features can be \emph{irrelevant}.
It has been observed that the existing algorithms for anomaly detection, in the context described above, lose their discriminating power very rapidly as even a small number of irrelevant features are added to the input vectors~\cite{Finke:2023ltw}.
In this paper, we will present approaches that address this issue within both the classifier-based and probability-density-based approaches to anomaly detection.

The algorithms on which we focus here are based on Boosted Decision Trees (BDTs), rather than neural networks.
BDTs tend to outperform neural networks on tabular data, where they can take advantage of the preferred basis implied by the input features~\cite{grinsztajn2022treebased,DBLP:journals/corr/abs-2110-01889}.
Unlike neural networks, which act like a black box, BDTs allow for interpretability of the model and feature importance.
Additionally, BDTs typically require less data preprocessing and feature engineering compared to neural networks for comparable dataset sizes.

The rest of the paper is organized as follows.
In \cref{sec:data}, we describe the ``signal'' and ``background'' data sets that are used in our analysis, and specify how we model the extraneous irrelevant features.
In \cref{sec:cwola}, we present a BDT-based classifier which uses the CWoLa approach to aid anomaly detection.
We show that before irrelevant features are added, the BDT algorithm achieves performance similar to that of NN-based classifiers.
However unlike the NN, the BDT performance does not deteriorate significantly when irrelevant features are present.
In \cref{sec:denest}, we show how the BDT can be used as a probability density estimator, providing a powerful tool for anomaly detection even when relevant features are correlated with $m_{JJ}$.
Furthermore, this algorithm is also robust in the presence of irrelevant features.
\cref{sec:conc} contains our conclusions. Technical details related to tuning of hyperparameters of the BDT algorithms are presented in \cref{app:tune,app:BDTpar}, while a case study of our methods' performance on a dataset with mutually dependent irrelevant features is discussed in \cref{app:mutual}.

In all plots in this paper, the curves showing performance of neural network anomaly-detection tools are generated using code provided at \url{https://github.com/HEPML-AnomalyDetection/CATHODE}.

\section{Dataset}
\label{sec:data}

The signal and background events used in this study are from the LHC Olympics 2020 R\&D dataset~\cite{gregor_kasieczka_2019_6466204}.
In particular, the SM background corresponds to QCD dijet events while the anomalous signal we want to detect is produced by the decay $W' \to X \qty(\to qq) Y \qty(\to qq)$.
Here $W'$, $X$ and $Y$ are hypothetical new bosons with masses \SI{3.5}{\TeV}, \SI{500}{\GeV} and \SI{100}{\GeV} respectively.
All events are produced using the \texttt{Pythia8}~\cite{Bierlich:2022pfr} and \texttt{Delphes 3.4.1}~\cite{deFavereau:2013fsa} Monte Carlo generators, and jets in each event are identified using \texttt{FastJet}~\cite{Cacciari:2011ma} using anti-$k_T$ clustering with $R = 1$.

The training (plus validation) set is constructed by combining \num{1000} randomly selected signal events with a sample of \num{1000000} background events.
For evaluation purposes, a separate test set is constructed by having \num{20000} signal events and \num{40000} background events, all of which lie inside the signal region (defined below).
This test set is not used during training.

The physically motivated relevant features are based on the two highest $p_T$ jets.
They include
\begin{itemize}
    \item $m_{JJ}$: invariant mass of the two jets, which will be the resonant feature.
    \item $m_{J_1}$: invariant mass of the lighter jet.
    \item $\Delta{m}_J$: absolute mass difference between the two jets' invariant masses.
    \item $\tau_{21}^{J_1}$, $\tau_{21}^{J_2}$: $n$-subjettiness ratios~\cite{Thaler:2010tr,Thaler:2011gf} of the two jets, defined by $\tau_{21} \equiv \tau_2 / \tau_1$.
\end{itemize}

Following~\cite{Collins:2019jip, Nachman:2020lpy, Hallin:2021wme}, we define the signal region (SR) by $m_{JJ} \in [3.3, 3.7] $ \si{\TeV}, and the sideband region (SB) by  $m_{JJ} \not\in [3.3, 3.7]$ \si{TeV}.
Additionally, for the CWoLa method, we also define a \emph{short side-band} (SSB) region, which extends to both sides of the SR by \SI{200}{\GeV}: $m_{JJ} \in \qty([3.1, 3.3] \cup [3.7, 3.9])$ \si{\TeV}.
These definitions will be used throughout the rest of the paper.

However, for a model-agnostic search, one would not know \emph{a priori} that the observables above are the only features of interest and would likely not have any principled way of excluding additional superfluous features.
To simulate such a scenario, we artificially augment the original dataset with features drawn from Gaussian distributions, which will be considered as our irrelevant features.
We vary the number of such irrelevant features and examine how much effect they have on anomaly detection performance.

\subsection{What Do We Mean by Irrelevant?}
\label{sec:irrel_def}

Even though the notion of an irrelevant feature is intuitively clear, it is necessary for us to define it more precisely.
We provide here two possible characterizations of ignorable irrelevant features, each suited to the respective anomaly detection method considered in the text.\footnote{The general definition of irrelevancy has been explored in \cite{irrelevant_features_definition}. The conditions stated below are less general, but suffice in the context of anomaly detection.}
We emphasize the ignorable aspect because features contributing no statistical power to the discrimination of hypotheses can nonetheless violate the assumptions of a particular analysis, leading to an expected degradation in performance.
We are interested in features that should not matter in the limit of an infinite amount of data.
\begin{itemize}
    \item CWoLa method:
    
    A feature $y$ is irrelevant if $p(m_{JJ} \in \text{SR} | y) = p(m \in \text{SR})$.
    
    \item Probability density estimation-based method:
    
    A feature $y$ is irrelevant if it is statistically independent of $m_{JJ}$ and the auxiliary (relevant) features: 
    $p(m_{JJ}, x_1, \dotsc, x_K, y) = p(m_{JJ}, x_1, \dotsc, x_K) \, p(y)$. This must hold in both the background and signal samples. Moreover, $p(y)$ in the signal and background samples must be identical. 
\end{itemize}

In this paper, we will explore the performance of anomaly detection algorithms as a function of the number of irrelevant features $N$.
As a baseline model, throughout this paper we assume that the irrelevant features $y_i$ are distributed according to a direct product of Gaussians:
\begin{equation} \label{eq:ifeatdist}
  p(y_i) = \prod_{i=1}^N \frac{1}{\sqrt{2\pi}} e^{-y_i^2/2}.    
\end{equation}
A vector of $N$ features drawn from this distribution is then tacked onto each event in the LHCO 2020 dataset described above, with no distinction made between signal and background events.
The features $y_i$ in the resulting dataset satisfy both of the irrelevancy definitions above. 

Within our baseline model, the $y_i$'s are mutually statistically independent among themselves.
This feature is not generic, and is not expected to always hold in realistic physics scenarios.
In \cref{app:mutual} we show that our anomaly detection algorithms continue to perform well when the irrelevant features are mutually dependent.

\subsection{Performance Metric}

As is standard, we shall present performance comparisons between NNs and our proposed methods in terms of the \emph{significance improvement characteristic} (SIC) curve, which is obtained by plotting the significance improvement,
\begin{align}
\label{eq:defSIC}
  \text{SIC} = \frac{\epsilon_S}{\sqrt{\epsilon_B}} \,,
\end{align}
against $\epsilon_S$.
Here $\epsilon_S$ is the fraction of correctly identified signal events (true positive rate), and $\epsilon_B$ is the fraction of background events incorrectly identified as signals (false positive rate).
It should be emphasized here that SIC is a meaningful metric only when the analysis is statistics-limited and not systematics-limited, and the sample is background-dominated; we shall assume that this is the case.

\section{CWoLa on a Tree: Classifier BDTs}
\label{sec:cwola}

In this section we compare the performance of BDT-based and NN-based CWoLa methods in the presence of irrelevant features.\footnote{In~\cite{Finke:2023ltw}, similar comparisons are made in the context of \emph{idealized anomaly detection}, in which perfect understanding of background is assumed. This included a more detailed, physical model of irrelevant features, while we consider a more realistic measurement scenario. We hence view the two studies as naturally complimentary.}

The CWoLa hunting~\cite{Collins:2018epr,Collins:2019jip} method attempts to construct the Neyman--Pearson optimal discriminator~\cite{Neyman:1933wgr} between a signal and a background where the signal is assumed to be dominantly present in the SR.
The key observation underlying this is that if the SSB and SR have different admixtures of signal and background, then the optimal signal-background discriminator is monotonically related to the optimal SSB-SR classifier and finding one produces the other, provided that the auxiliary features $\vec{x}$ are independent of the resonant mass $m_{JJ}$ for the background.
While this is a theoretical guarantee, finding an optimal SSB-SR classifier can be difficult in practice.
This is because at very low $S/B$ ratio, SSB and SR events largely overlap in feature space with very similar distributions, and most modern machine learning models are flexible enough to mistake local fluctuations for actual excess of signal events (\emph{i.e.}, over-fitting).
This situation is particularly exacerbated in the presence of irrelevant features, because they provide additional sources of statistical fluctuations in a higher dimensional space.

The above consideration do not actually select for a method of approximating the SSB-SR classifier.
In studies involving the CWoLa hunting method, the classifier typically consists of a fully-connected feedforward neural network.
However, it is well-known that neural networks do not fare well with irrelevant inputs, and this is especially so when they are applied in the CWoLa setting for reasons above.

On the other hand, tree-based models are known to be innately robust against irrelevant features~\cite{hastie01statisticallearning, grinsztajn2022treebased}, an observation usually attributed to the way they are constructed --- for most tree-based models they are built by performing cuts in feature space to greedily minimize metrics such as information gain, meaning that they already have some degree of internal feature selection built in.
Here we capitalize on this empirical observation and apply BDT-based CWoLa to a more realistic setting where inevitably there will be a lot of irrelevant features.

In what follows, we use \texttt{xgboost} as a reference BDT model to compare with a fully-connected feed-forward NN. \texttt{xgboost} is chosen since it is widely considered as (one of) the state-of-the-art gradient boosting tree algorithms in terms of speed and accuracy. For detailed descriptions of the \texttt{xgboost} algorithm, refer to~\cite{Chen_2016}.

\subsection{Training Procedures}

\begin{table}
    \centering
    \begin{tabular}{c|c|c|c|c|c}
         \texttt{n\_estimators} & \texttt{max\_depth} & \texttt{eta} & \texttt{alpha} &  \texttt{lambda} & \texttt{subsample} \\
    \hline
         292 & 9 & \num{6.2e-3} & 50 & 74 & 0.75 \\
    \hline
    \end{tabular}
    \caption{One set of \texttt{xgboost} hyperparameters found for the dataset without any irrelevant features. We use default values for other hyperparameters. Refer to \cref{app:tune} or~\cite{Chen_2016} for a more detailed discussion of these hyperparameters.}
    \label{tab:xgbparam}
\end{table}

For training the CWoLa classifiers\footnote{The data handling and training procedures are the same as in~\cite{Hallin:2021wme}.
Here we summarize them for the sake of completeness.}, we select from the raw training set events for which $m_{JJ} \in [3.1, 3.9]$ \si{TeV}.
This results in roughly \num{250000} training events with about 760 signal events in the SR, which corresponds to $S/B \approx 0.6\%$ and $S/\sqrt{B} \approx 2.2$.
Classifiers are then trained to differentiate between SR and SSB labels.

The NN-based classifier is constructed by a fully-connected feed-forward neural network with 3 hidden layers, each of which has 64 neurons.
The network is trained for \num{100} epochs with binary cross-entropy loss using the \texttt{Adam} optimizer~\cite{KingBa15} and learning rate set to \num{e-3}.
During training, only half of the dataset constructed above is used for actual training while the other half is used for validation purposes.
In particular, the 10 epochs with the lowest validation error are used to construct an ensemble of 10 classifiers. 

\begin{figure}
    \centering
    \begin{subfigure}{0.49\textwidth}
        \centering
        \includegraphics[width=\textwidth]{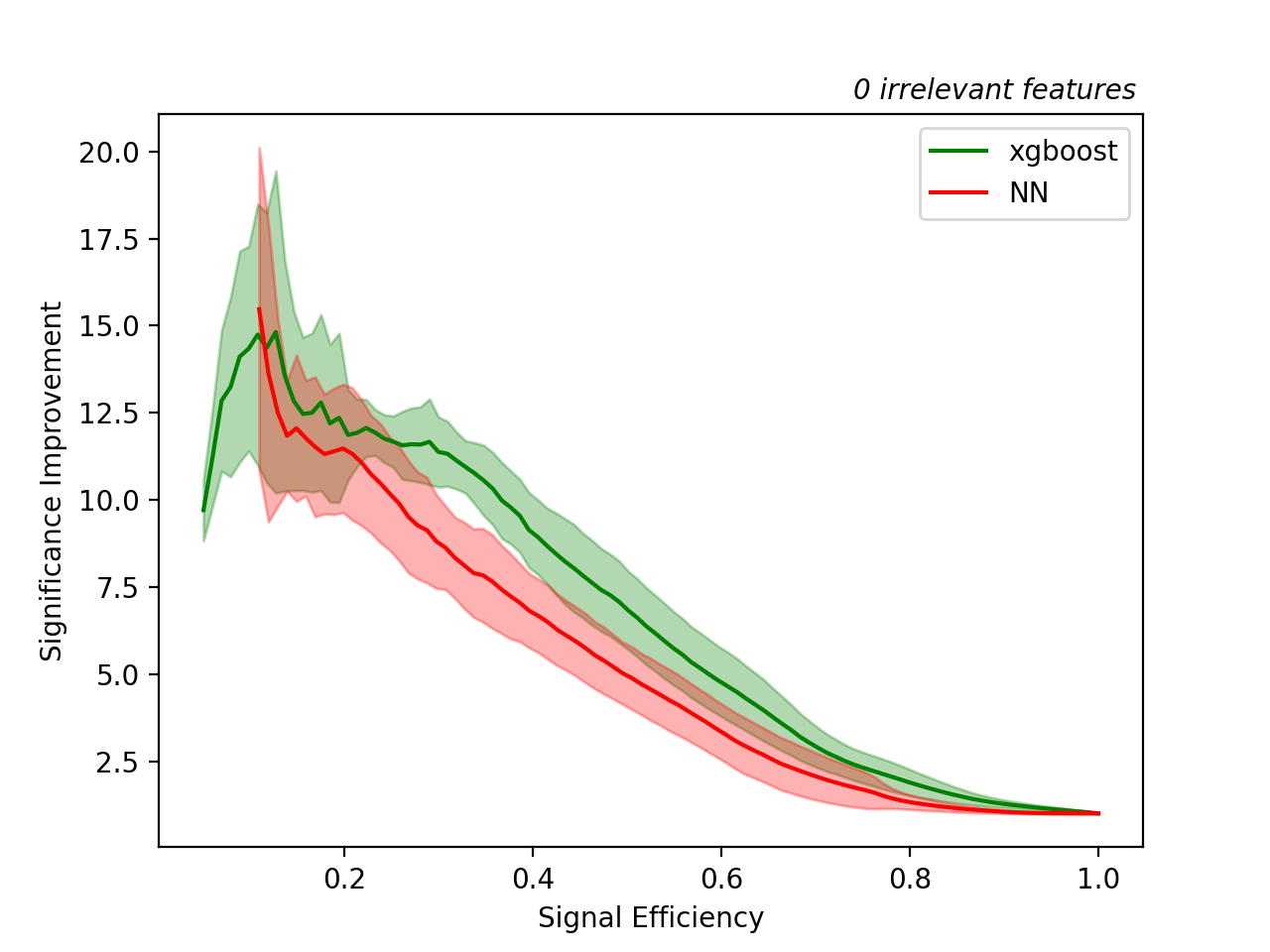}
    \end{subfigure}
    \hfill
    \begin{subfigure}{0.49\textwidth}
        \centering
        \includegraphics[width=\textwidth]{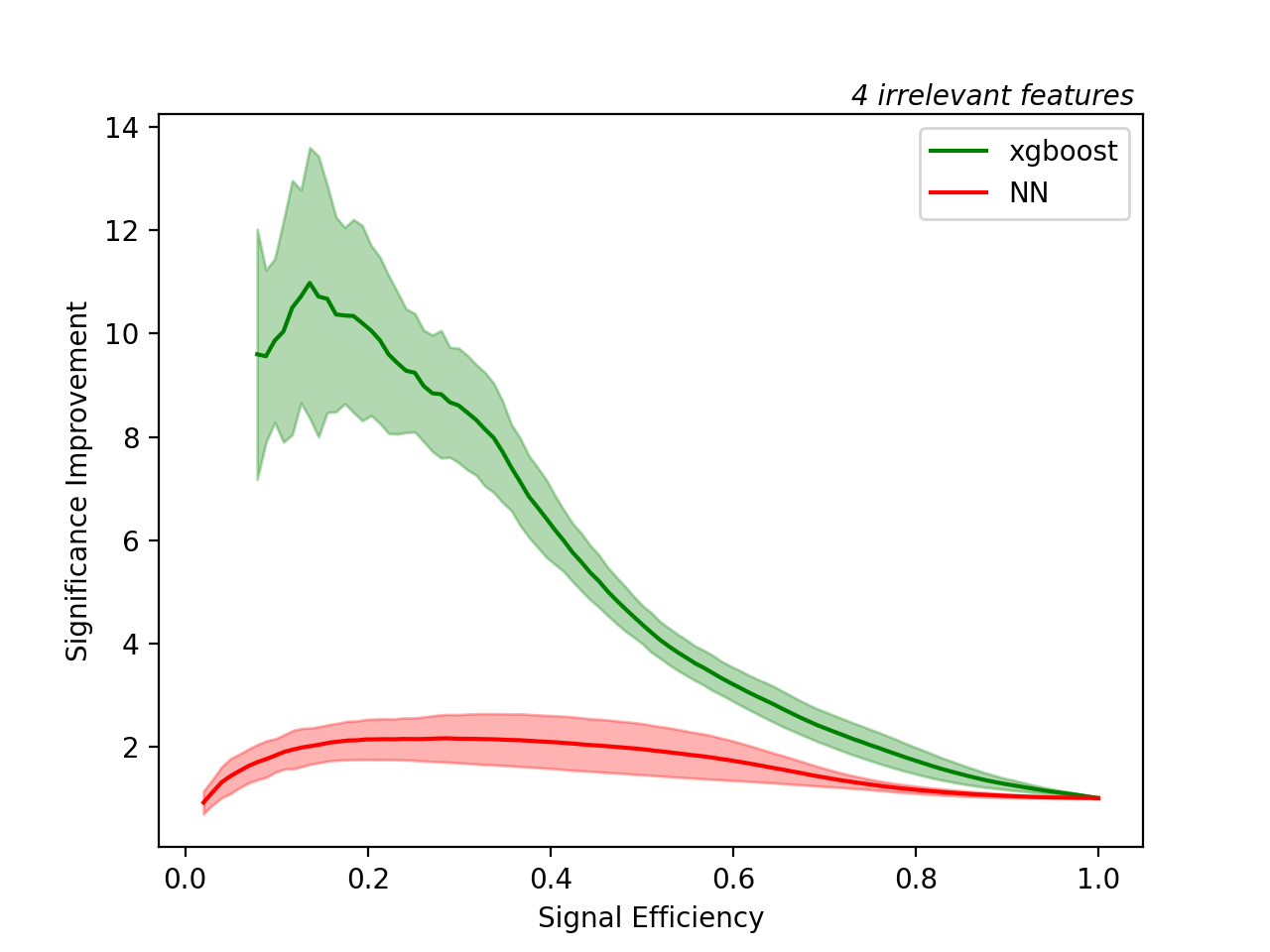}
    \end{subfigure}
    \\
    \begin{subfigure}{0.5\textwidth}
        \centering
        \includegraphics[width=\textwidth]{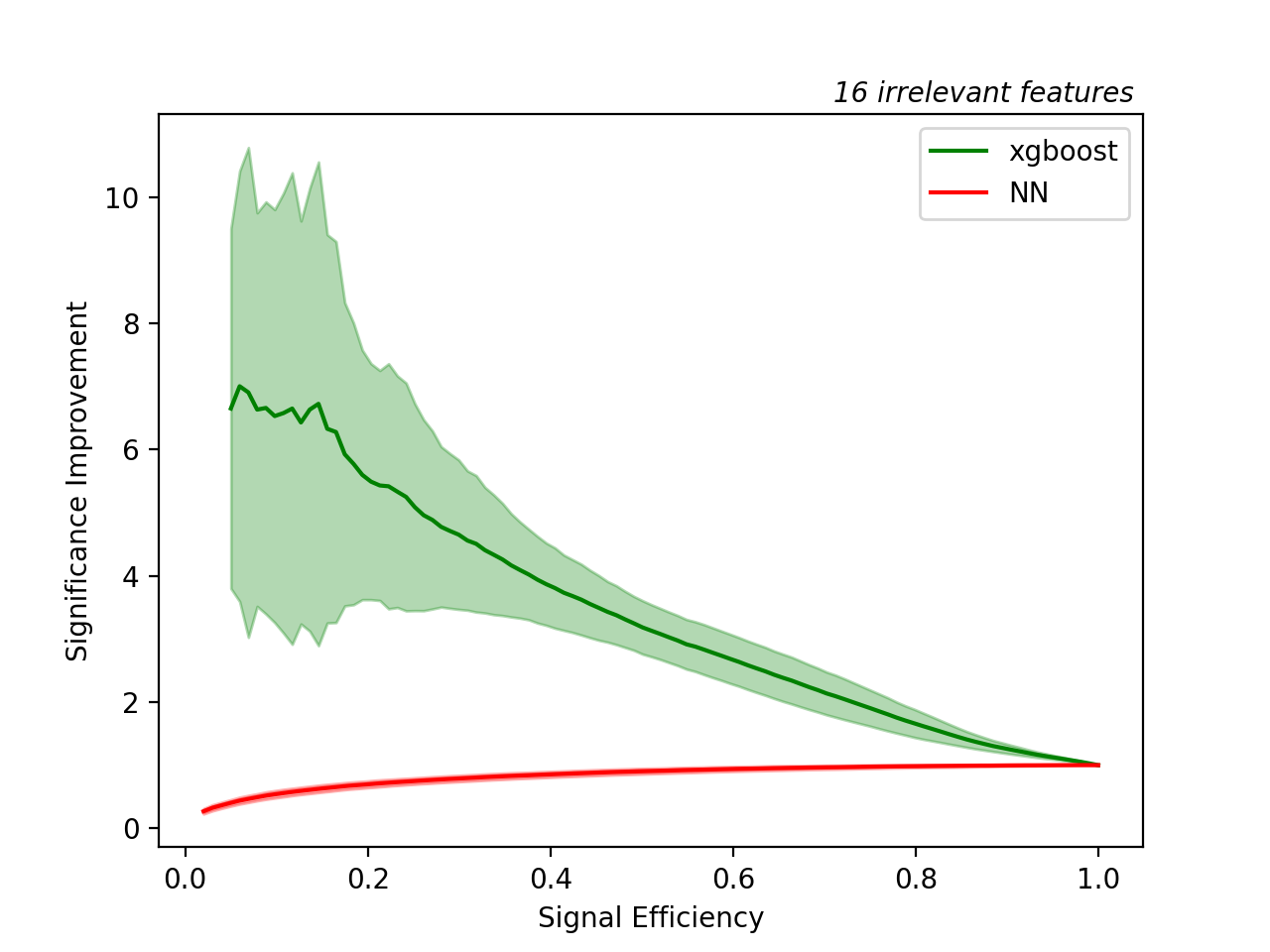}
    \end{subfigure}
    \caption{Performance comparisons between BDT-based (green) and NN-based (red) CWoLa methods for dataset augmented with 0, 4, and 16 irrelevant Gaussian features. The same \texttt{xgboost} hyperparameters are used to train in all 3 cases. The solid lines represent average SIC value across a classifier ensemble defined in the text, and the bands refer to 1 standard deviation of SIC. It is important to note that for neural networks, the bands correspond to variability for a \emph{fixed} set of hyperparameters, while for \texttt{xgboost} they correspond to variability across \emph{different} hyperparameters found by Bayesian optimization. Clearly, \texttt{xgboost} is far more robust against the inclusion of irrelevant features than neural networks.}
    \label{fig:SIC_cwola}
\end{figure}

For the \texttt{xgboost}-based classifier, we employ a 10-fold cross-validation so that the entire training set is utilized during actual training.
Specifically, we use the cross-validation process to tune \texttt{xgboost}'s hyperparameters to the dataset without irrelevant features\footnote{This is necessary because the default hyperparameters are far too aggressive and lead to severe overfitting.}.
Details of this procedure can be found in \cref{app:tune}.
Since the hyperparameter optimization procedure is stochastic, we find 10 independent sets of hyperparameters, each of which is used to train a separate classifier and they together form an ensemble of 10 classifiers.
The same set of hyperparameters is also used to train dataset augmented with irrelevant features.
In \cref{tab:xgbparam}, we show one set of hyperparameters found.

\subsection{Performance Comparison}

\begin{figure}
    \centering
    \begin{subfigure}{0.49\textwidth}
        \centering
        \includegraphics[width=\textwidth]{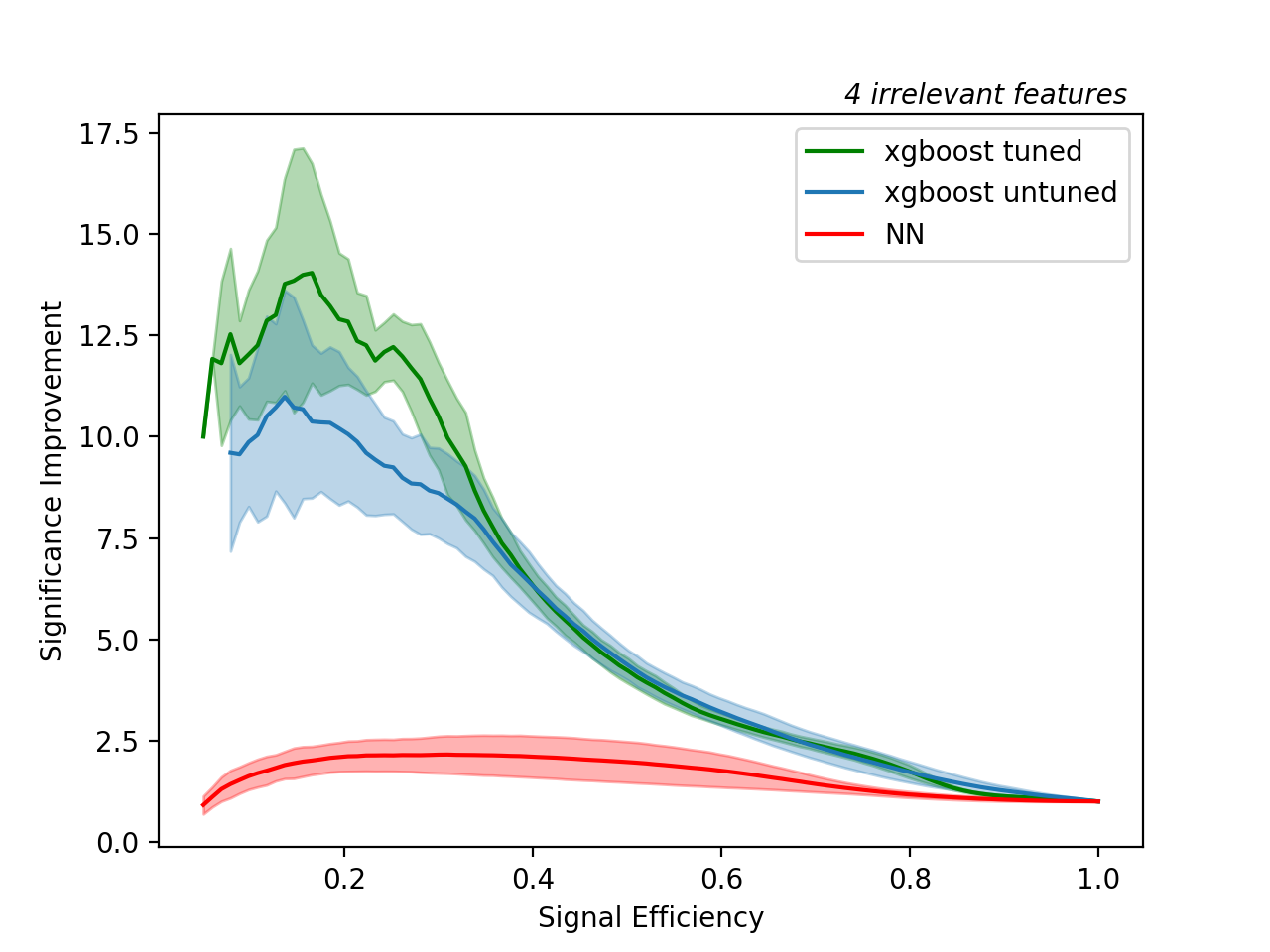}
    \end{subfigure}
    \hfill
    \begin{subfigure}{0.49\textwidth}
        \centering
        \includegraphics[width=\textwidth]{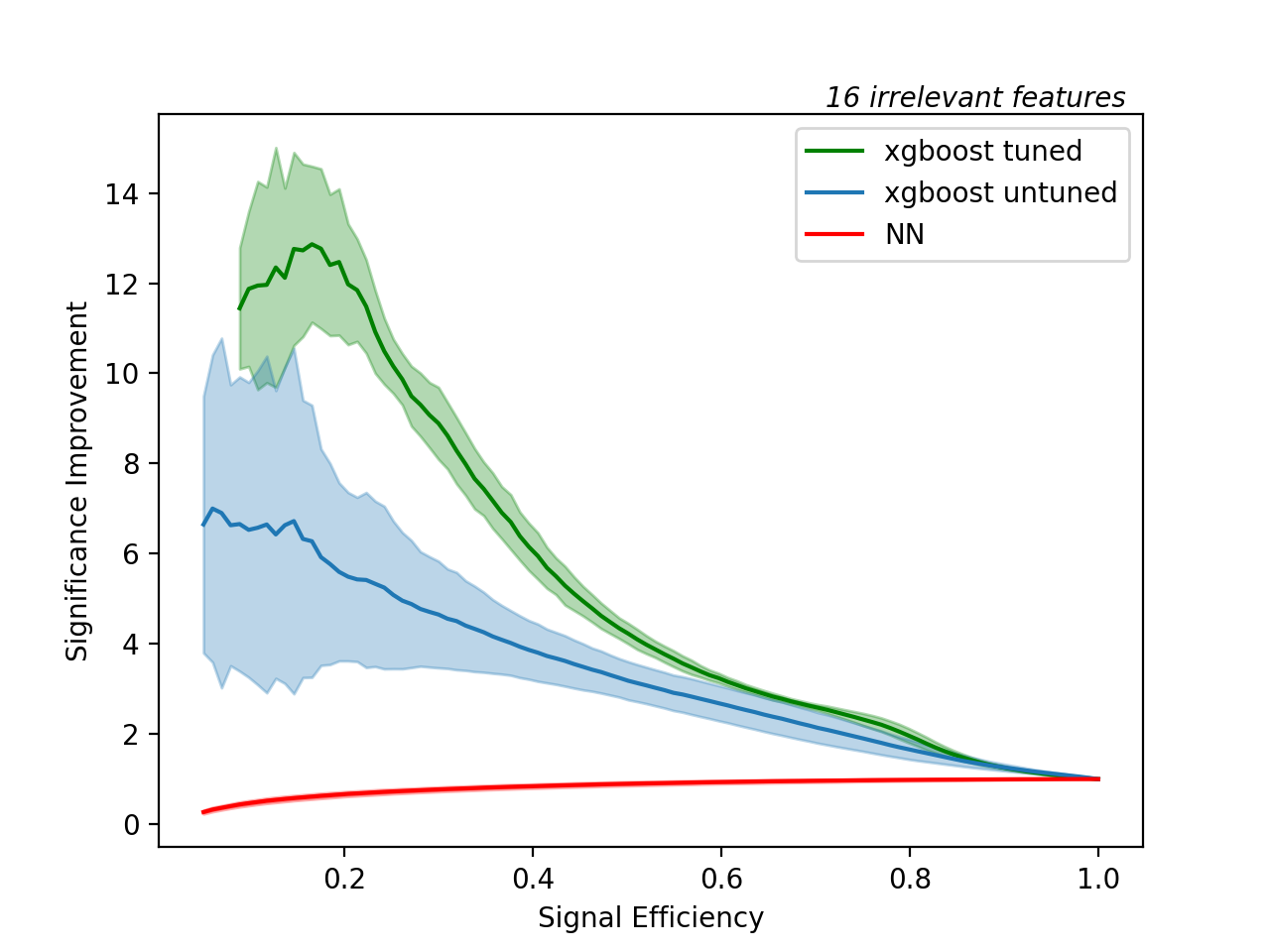}
    \end{subfigure}
    \caption{SIC curves for the \texttt{xgboost} classifier with hyperparameters optimized for the dataset with no irrelevant features (blue), and the same classifier with hyperparameters re-optimized each time more irrelevant features are added (green). With proper tuning, much of the original performance can be recovered even when the dataset has a large fraction of irrelevant features. NN-based classifier's performance on the same datasets (red) is provided to help guide the eye.}
    \label{fig:SIC_cwola_tuned}
\end{figure}

\begin{figure}
    \centering
    \includegraphics[width=\textwidth]{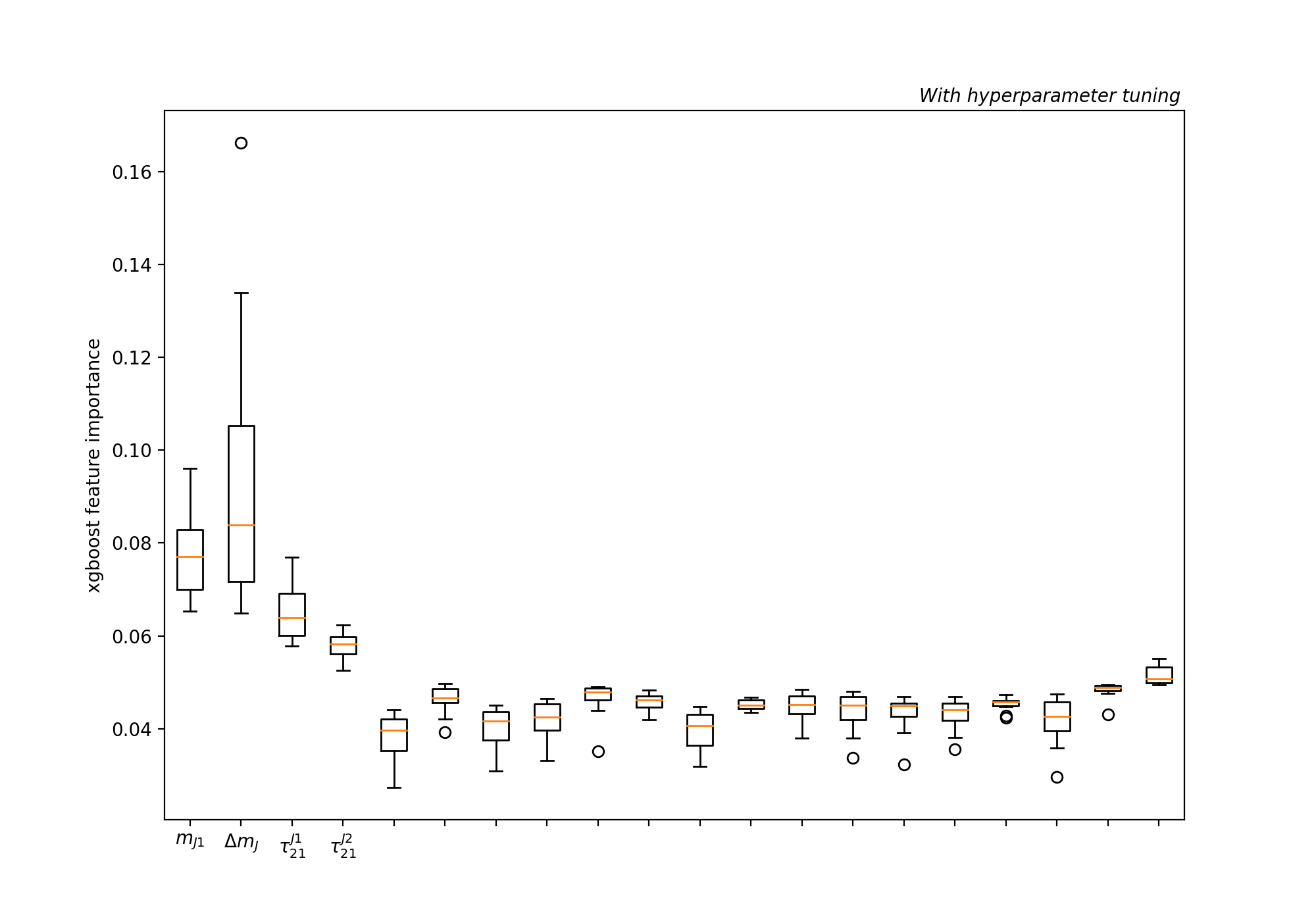}
    \caption{Box plot of feature importance values by 10 independent \texttt{xgboost} classifiers (with properly tuned hyperparameters) in the case with 16 irrelevant features. The unlabelled boxes correspond to the artificially introduced Gaussian noise features. The relevant features in the original dataset are found by the method to be more important for signal/background discrimination compared to the Gaussian noise.}
    \label{fig:featimp}
\end{figure}

The performances of \texttt{xgboost}-based and NN-based CWoLa are shown in \cref{fig:SIC_cwola}.
Both \texttt{xgboost}-based and conventional CWoLa perform similarly in the absence of irrelevant features.
However, when irrelevant features are present, the performance degradation of the neural network is much more severe than that of the BDT.
In particular, in the regime of large number of irrelevant features (relative to number of relevant ones), the neural network-based CWoLa method becomes essentially ineffective.
On the other hand, while BDTs-based CWoLa (without dedicated hyperparameter tuning) also suffers from the presence of irrelevant features, it is far more resilient.
In particular, even with 16 noisy features, the classifier can still attain an average maximum significance improvement of around 7.
This performance can be further improved by dedicated hyperparameter optimization each time more irrelevant features are added.
The performances of \texttt{xgboost} on the augmented dataset when the hyperparameters are properly tuned are shown in \cref{fig:SIC_cwola_tuned}.
Impressively, much of the model's original performance in the absence of irrelevant features can be recovered without too much of computational burden (relative to neural networks).
This shows the overall superiority of using BDTs when the input data is of tabular form in the context of CWoLa hunting.

Another added bonus of using a tree-based classifier is that there exists a naturally defined and easily computable notion of \emph{feature importance}~\cite{hastie01statisticallearning}.
Recall how a tree-based model is constructed: cuts along different feature directions are selected so as to greedily minimize the loss function.
Hence, for each feature one can compute how much it contributes to the overall decrease in loss.
This adds a layer of interpretability to the model which can potentially be used to shed light on what features are more relevant in discerning signal from background.\footnote{It is important to emphasize that this is meaningful only when the features are mostly uncorrelated from each other. If not, it becomes difficult to isolate the effect of each individual feature.}

We can use this notion of feature importance to understand why \texttt{xgboost} is so much more robust compared to neural networks.
In \cref{fig:featimp}, we show a box plot of feature importance values as given by the 10 different classifiers in the ensemble in the case of having 16 irrelevant features.
Strikingly, the model clearly utilizes the relevant features much more than the irrelevant ones, corroborating with our intuition that tree-based models by nature perform a certain degree of internal feature selection.
This is likely the reason why the \texttt{xgboost}-based CWoLa shows such favorable results.

In conclusion, even a naive direct application of BDT algorithms to CWoLa method can significantly increase its robustness to irrelevant features compared to NN-based CWoLa. 

\section{Probability Density Estimation with BDTs}
\label{sec:denest}

Even though the CWoLa method can achieve significant sensitivity improvement in anomaly detection, its success hinges on the independence of the auxiliary features $\vec{x}$ with $m_{JJ}$ under the background hypothesis, which is quite a strong assumption and does not hold in general.
When this assumption is sufficiently violated, CWoLa performance drops drastically~\cite{Nachman:2020lpy, Hallin:2021wme}.

Since this is a strong assumption the does not always hold in physical analyses scenarios of interest, various methods have been proposed to circumvent it.
In particular, we examine the anomaly detection with density estimation (ANODE) method~\cite{Nachman:2020lpy}, which was originally implemented using normalizing flows.
While this is not the state-of-the-art anomaly detection method, it is chosen since the lessons learned here can be easily transferred to other similar density-estimation-based methods.

Unlike CWoLa, the ANODE method tries to estimate the two probability densities: $p(\vec{x} | m)$ of the full data set, and $p(\vec{x} | m, \text{bkgd})$ of the background only (estimated from the sideband regions and extrapolated in the signal region).
Then, the likelihood ratio
\begin{align} \label{eq:anodeR}
    R = \frac{p(\vec{x} | m)}{p(\vec{x} | m, \text{bkgd})}
\end{align}
is computed in the signal region.
This ratio can be shown to be optimal (in the Neyman--Pearson sense) without any need of additional assumptions.

In other words, the ANODE method mainly consists of two steps:
\begin{itemize}
    \item Estimate the full density $p(\vec{x}|m)$ directly from data,
    \item Estimate the background density $p(\vec{x}| m, \text{bkgd})$ by interpolating from the SB regions into the SR region.
\end{itemize}
Note that a hidden assumption here is that the auxiliary features have smooth distributions over the SR in the background, for otherwise there would be no reason to believe that interpolation would give a sensible background estimate.
This is often true in practice given that the SR is rather small.

Below we explain how the same steps can be achieved using boosted trees.

\subsection{Boosted Density Estimation Trees}

Motivated by the success of using BDTs with the CWoLa method, here we examine the possibility of applying them to density estimation.
Specifically, we follow the tree density estimation algorithm presented in \cite{awaya2023unsupervised}, which we describe briefly here. For details, please refer to the original literature.

Conceptually, the BDT density estimation algorithm is very similar to that of normalizing flows~\cite{Papamakarios_NF} --- they both model the transformation between the target density and some base density as a composition of simple, bijective maps.
Importantly, each composition is thought of as a round of boosting just as in the traditional algorithm.

The major difference between the two is that in the case of BDT, the transformations are built from cuts in the feature space (selected so that they locally minimize the KL divergence between the empirical distribution of the transformed data and the base distribution, which is uniform in our case) with Jacobians admitting closed-form evaluations, whereas for normalizing flows they are typically parameterized by neural networks~\cite{Papamakarios_NF}.
The density estimated by the corresponding tree is then a leaf-wise constant function. After each round of boosting, one can define and compute the difference between the learned density and the target density, which is used as the target density for next round.
This procedure is recursively performed until some termination condition is satisfied.

\subsubsection{Copula}

When estimating probability densities, it is often helpful to separate the task of estimating marginal densities and from the task of estimating the dependence structure between variables.
This can be achieved explicitly by Sklar's theorem~\cite{Skla59}, which states, as part of the theorem, that any multivariate probability density $p(x_1, \dotsc, x_d)$ (satisfying some very mild conditions) can be represented in the following form:
\begin{align} \label{eq:copula}
  \begin{split}
    p(x_1, \dotsc, x_d) &= c\big(F_1(x_1), \dotsc, F_d(x_d)\big)
                           f_1(x_1) \dotsm f_d(x_d) \\
                        &\equiv \tilde{c}(x_1, \dotsc, x_d)
                                 f_1(x_1) \dotsm f_d(x_d) \,,
  \end{split}
\end{align}
where the $f_i$'s are the marginal densities, the $F_i$'s are the corresponding cumulative distribution function (CDF), and $c$ is the so-called copula density function.
This copula function completely encapsulates the information about dependence structure among variables.

The input data considered in this paper consists of the dijet mass $m_{JJ}$, the auxiliary features $x_1, \dotsc, x_K$, and the additional features $y_1\ldots y_N$ which contain no information relevant for anomaly detection.
Moreover, we assume that irrelevant features are statistically independent of $(m_{JJ}, x_1, \dotsc, x_K)$.
With this assumption, the copula decomposition takes the form 
\begin{multline}\label{eqn:factorization}
   p(m_{JJ}, x_1, \dotsc, x_K, y_1, \dotsc, y_N)
     = \tilde{c}(m_{JJ}, x_1, \dotsc, x_K) \,
         \tilde{c}( y_1, \dotsc, y_N) \\
           \times f_1(x_1) \dotsm f_K(x_K) g_1(y_1) \dotsm g_N(y_N)\,.   
\end{multline}
Furthermore, if the irrelevant features are mutually independent among themselves, as in \cref{eq:ifeatdist}, the corresponding copula function is trivial, 
\begin{equation}\label{eqn:triviality}
\tilde{c}\qty( y_1, \dotsc, y_N)=1.
\end{equation}
Then, the likelihood ratio in \cref{eq:anodeR} takes the form
\begin{align} \label{eq:anodeRcop}
    R = \frac{\tilde{c}(m, x_1, \dotsc, x_K)}
             {\tilde{c}(m, x_1, \dotsc, x_K | \text{bkgd})}
          \prod_{k=1}^K \frac{f(x_k)}{f(x_k | \text{bkgd})} \,.
\end{align}
Note that by using the copula decomposition, the dependence on the irrelevant features in $R$ drops out in both the marginal and the copula densities.
The cancellation in the marginal density ratio is easy to ensure in practice since it simply relies on univariate density estimation.
As for the copula density, the model needs to be able to learn that it is independent of $(y_1\ldots y_N)$.
This is where the tree-structure shines --- similar to the supervised case, the tree model should be able to learn to not cut along the irrelevant directions, since they do not contribute much towards the decrease in KL divergence when estimating the copula density.

In view of the discussion above, we follow the basic two-stage strategy suggested in~\cite{awaya2023unsupervised}: we first fit models to the marginal variables, and then we use the learned CDF to transform them to the copula space on which we estimate the corresponding copula density.
The final learned density is given by \cref{eq:copula}.
Note that neither the copula factorization, \cref{eqn:factorization}, nor the mutual independence of the irrelevant features, \cref{eqn:triviality}, are hardwired into our algorithm.
Rather, these features are efficiently learned by the BDT from the structure of the training data.
The high quality of the trained tree model contributes to robustness of the anomaly detection algorithm in the presence of irrelevant features.
At the same time, the underlying BDT has sufficient flexibility to remain useful when the structure of the input data is more complex.
This is evidenced by the example with mutually dependent irrelevant features considered in \cref{app:mutual}.

\subsection{Interpolation}

Once the probability density in the SB region is estimated, the next step is to interpolate it into the SR.
Unlike the NN, a tree-based density estimator does not automatically provide such an interpolation, and it needs to be implemented by hand.
This represents an additional step in the algorithm, but has an inherent advantage of being controllable, in contrast to a black box-like interpolation performed by the NN. 
As a baseline, we employ a naive linear interpolation:
\begin{align} \label{eq:linterpol}
  p(\vec{x}|m) = p(\vec{x}|m_L)
                 + \frac{p(\vec{x}|m_R) - p(\vec{x}|m_L)}
                     {m_R - m_L} (m - m_L), \quad
  m \in (m_L, m_R)\,,
\end{align}
where $m\equiv m_{JJ}$; $m_L$ and $m_R$ are the lower and upper boundaries of the signal region in $m_{JJ}$; and the vector $\vec{x}$ includes both auxiliary (relevant) and irrelevant features.
While more elaborate methods of interpolation exist~\cite{Andreassen:2020nkr, Raine:2022hht, Sengupta:2023xqy}, this simple form is chosen here for the following reasons:
\begin{itemize}
    \item Under the assumption that the SR is sufficiently small and that the SB is not significantly signal-contaminated, we expect linear interpolation to give reasonable results (there are however some subtleties, see \cref{sec:correl}).
    \item More importantly, in \cref{eq:linterpol}, the interpolated density is explicitly linear in the learned density. In the ideal case that the irrelevant features' densities factorize from the learned density, this property ensures that dependence on irrelevant variables will be cancelled out in the construction of the likelihood ratio. As we shall see below, this linearity property is important in ensuring robustness.
\end{itemize}

\subsection{Training and Evaluation Procedures}

\begin{table}
  \centering
  \begin{tabular}{c|c|c|c|c}
    & \texttt{n\_estimators} & \texttt{max\_depth} & \texttt{lr} & \texttt{gamma} \\
    \hline
    marginal & 100  & 10 & 0.1 & 0.3 \\
    \hline
    copula   & 2500 & 50 & 0.1 & 0.3 \\
    \hline
  \end{tabular}
  \caption{Hyperparameters used to estimate the marginal and copula densities with the tree-based algorithm in~\cite{awaya2023unsupervised}. Please refer to the original literature or \cref{app:BDTpar} for meanings of these parameters.}
  \label{tab:BDThparams}
\end{table}

The NN-based density estimator we use in our comparison is a masked-autoregressive flow (MAF).
The training procedure for a MAF is the same as in \cite{Hallin:2021wme}, and we refer readers to the original paper for details. 

The training of the tree-based density estimator is done by feeding the algorithm the entire training dataset (\num{560000} events), with hyperparameters listed in \cref{tab:BDThparams}.
The performance of the BDT density estimation algorithm is fairly insensitive to the choice of hyperparameters as long as the resulting model is sufficiently expressive, due to the fact that we are  estimating the density from the background sample for which we have large statistics.
(This is in contrast with the CWoLa case, where the BDT needs to learn a small difference between the distributions of auxiliary features in the signal and side-band regions.)
The preprocessing of~\cite{Hallin:2021wme} is not necessary in this case, since trees are invariant under monotonic transformations.
The trained model is then used to evaluate $p(\vec{x} | m)$, $p(\vec{x} | m_L)$, and $p(\vec{x} | m_R)$.
The latter two are used to estimate the background density in the signal region according to \cref{eq:linterpol}.

The performance of each method is evaluated on a separate test set consisting of \num{20000} signal events and \num{60000} background events, all of which lie in the SR.
In particular, we train the tree-based model for 10 random training-validation-test splits to cross-validate its variance.
The same is not done for neural networks due to their high computational costs. Comparisons are shown in the next section.

\subsection{Performance Comparison}
\label{sec:perf}

\begin{figure}
    \centering
    \begin{subfigure}{0.49\textwidth}
        \centering
        \includegraphics[width=\textwidth]{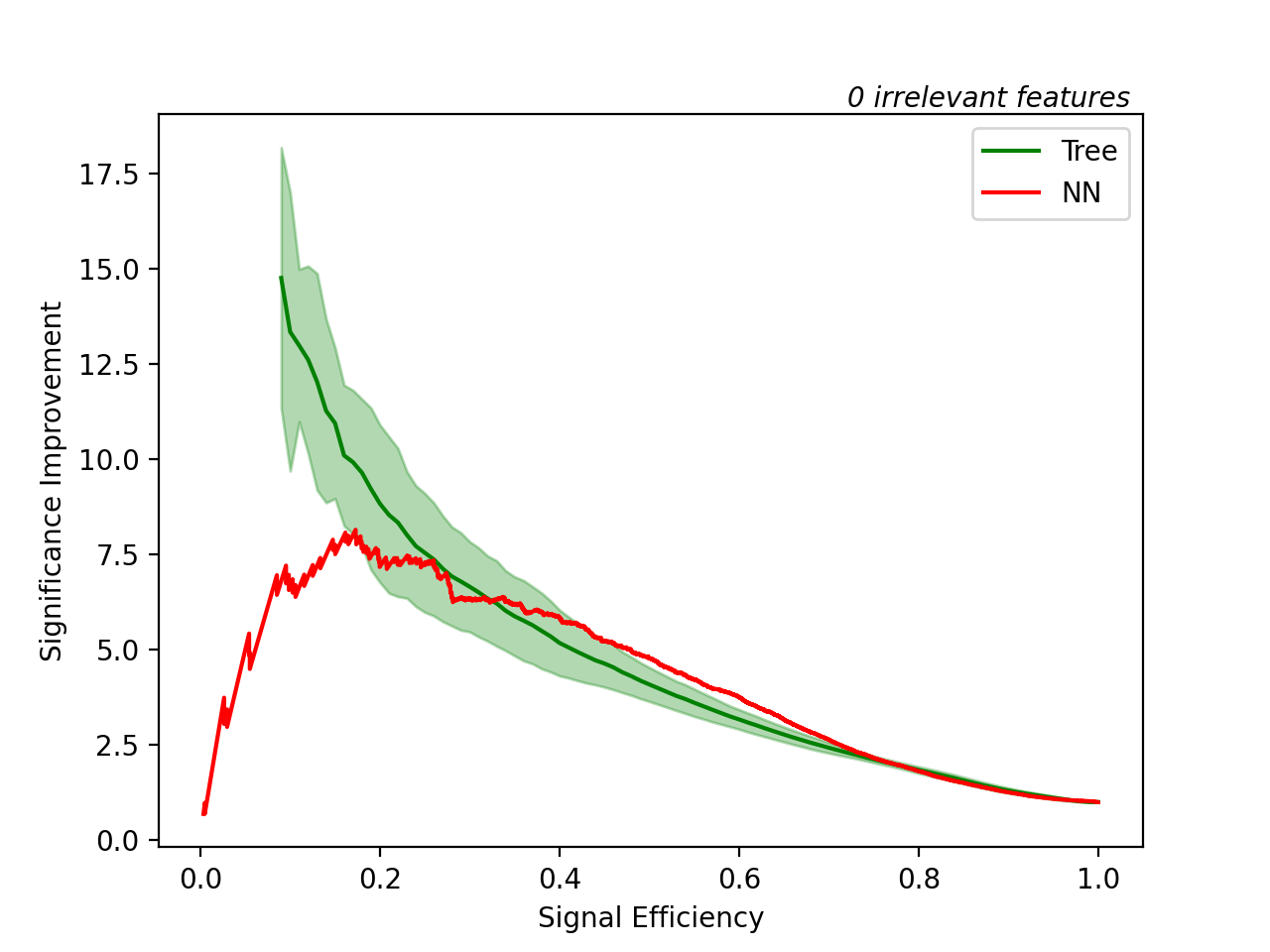}
    \end{subfigure}
    \hfill
    \begin{subfigure}{0.49\textwidth}
        \centering
        \includegraphics[width=\textwidth]{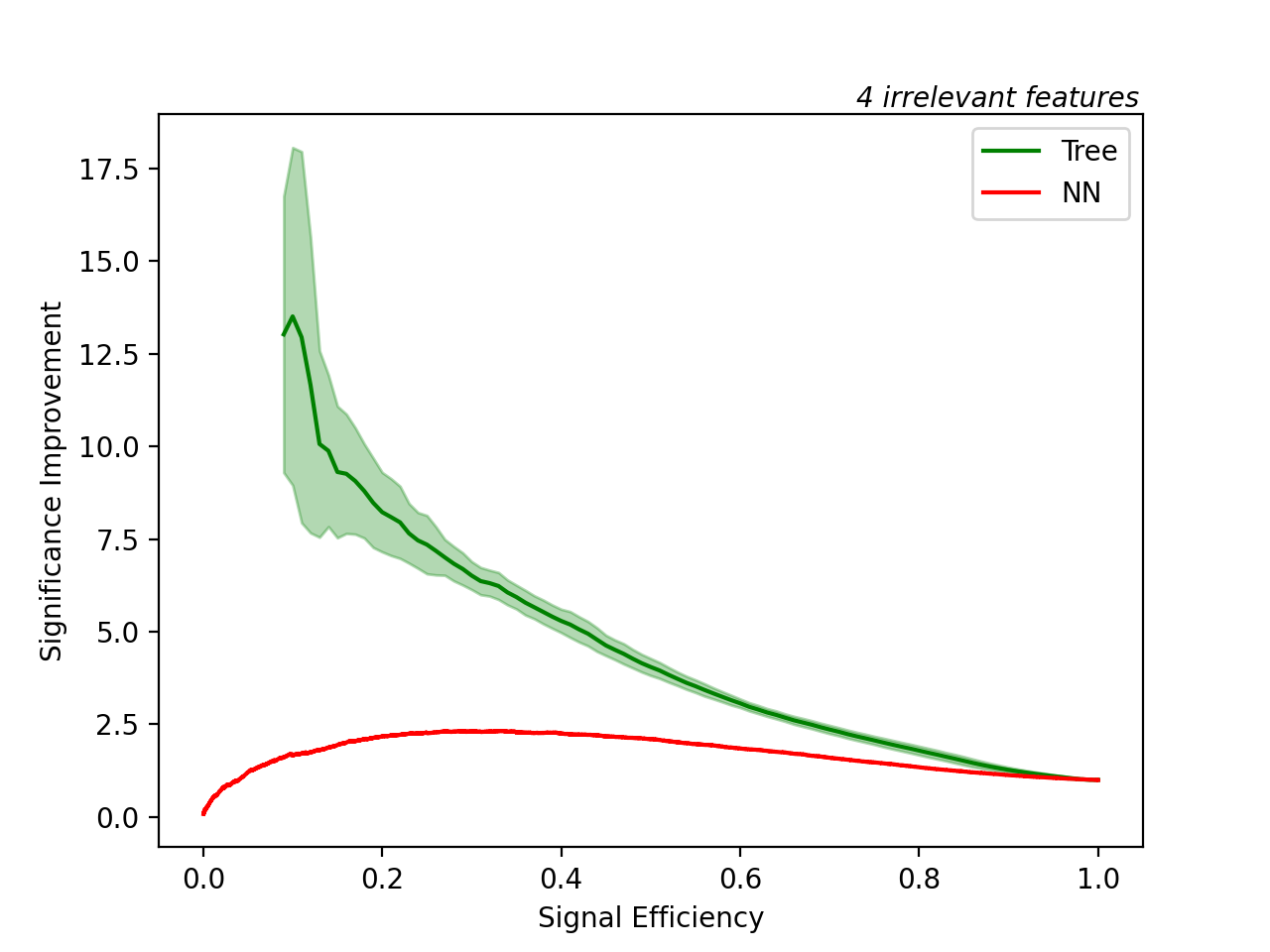}
    \end{subfigure}
    \\
    \begin{subfigure}{0.5\textwidth}
        \centering
        \includegraphics[width=\textwidth]{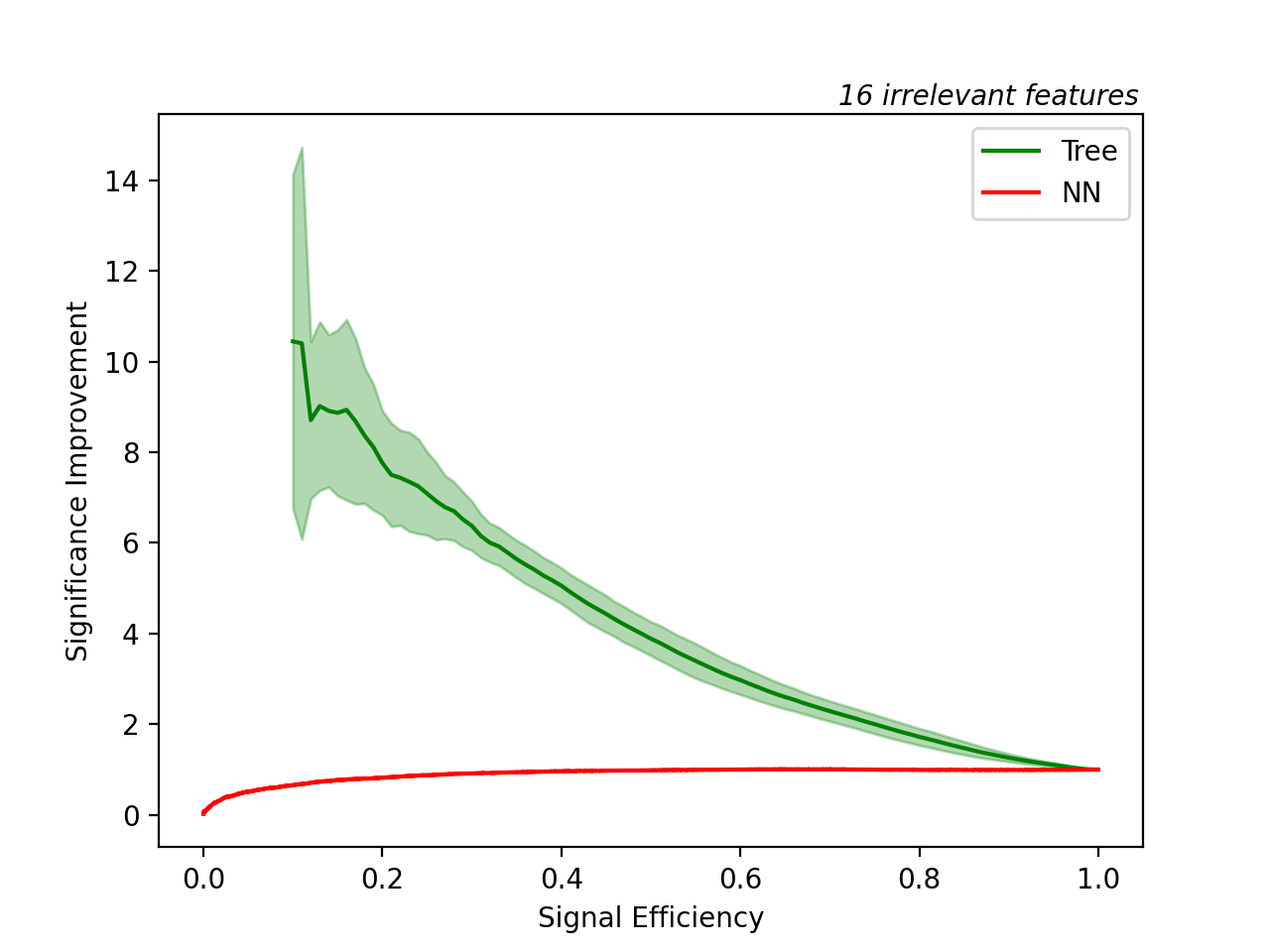}
    \end{subfigure}
    \caption{Performance comparison between the BDT implementation of density estimator (green) and the NN-based implementation found in~\cite{awaya2023unsupervised} (red), for the original LHCO dataset. The error bands show 1 standard deviation of significance improvement across 10 random training-validation-test splits. The BDT implementation shows superior performance both with and without inclusion of irrelevant features. In particular, even with 16 irrelevant features added, the BDT only shows a small level of degradation.}
    \label{fig:SIC_ANODE}
\end{figure}

In \cref{fig:SIC_ANODE}, we show the performance comparisons between MAF- and tree-based density estimation algorithms.
Without irrelevant features, the tree-based algorithm already provides a significant improvement over the NN in the low signal efficiency region.
Furthermore, just as in the case of CWoLa, the MAF-based algorithm suffers from severe performance degradation as the number of irrelevant features increases.
In the case where 16 irrelevant features are added, the method is essentially no different from a random classifier.
On the other hand, the tree-based algorithm is remarkably robust, showing almost no degradation of performance with up to 16 irrelevant features.

As an additional note, the success of the tree-based density estimation algorithm also shows that a simple linear interpolation for background estimation is very effective, at least for the LHCO dataset.
We believe this is evidence that more considerations should go into studying the interpolation method instead of relying on a black box like NNs.

\subsection{Correleated Auxiliary Features}
\label{sec:correl}

\begin{figure}
    \centering
    \begin{subfigure}{0.49\textwidth}
        \centering
        \includegraphics[width=\textwidth]{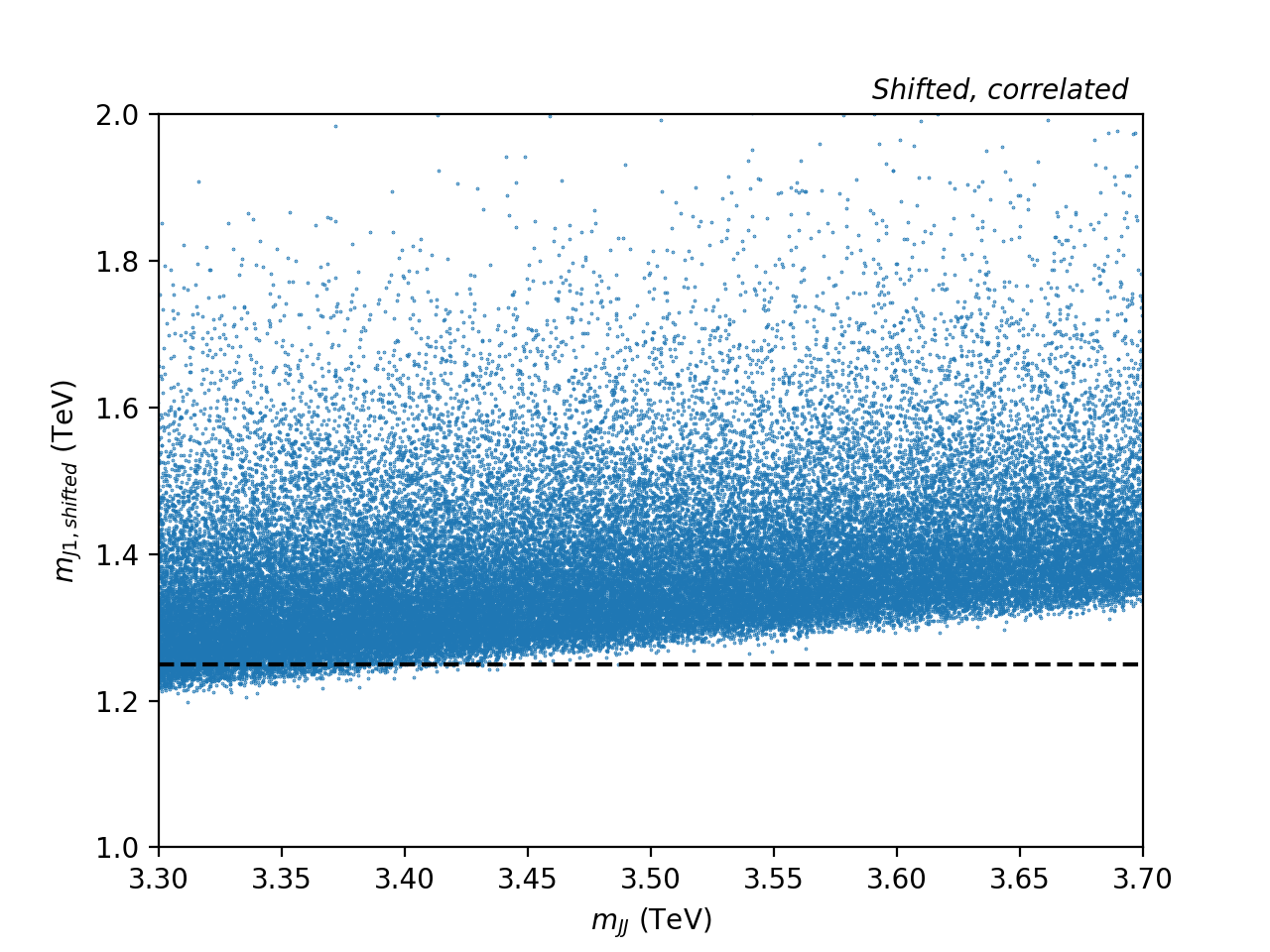}
    \end{subfigure}
    \hfill
    \begin{subfigure}{0.49\textwidth}
        \centering
        \includegraphics[width=\textwidth]{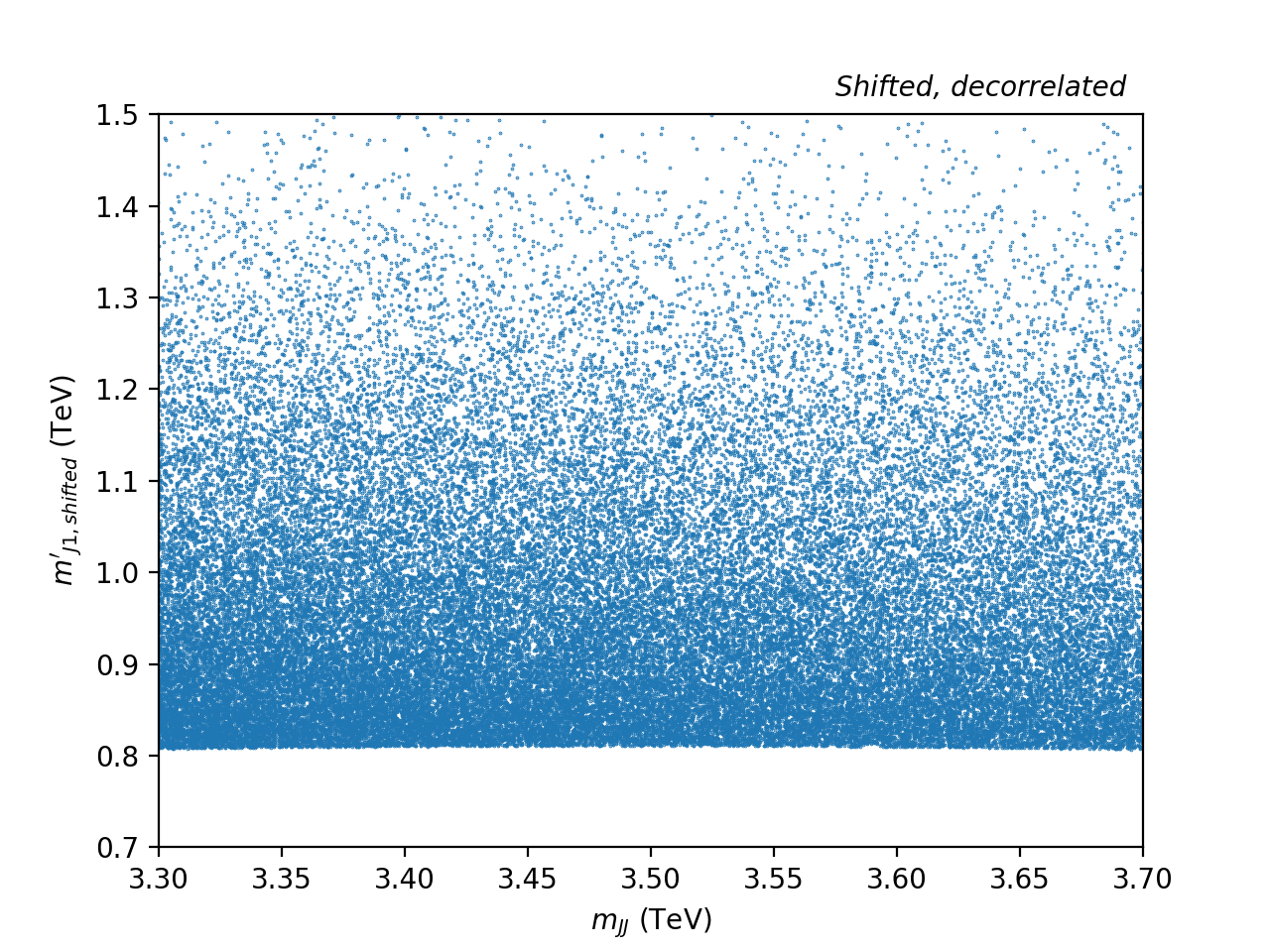}
    \end{subfigure}
    \caption{Scatter plot of $m_{J1}$ against $m_{JJ}$ for the log-shifted dataset over the signal region. Before decorrelation (left panel), the plot clearly shows why a naive linear interpolation should fail 
    --- the interpolation over the dashed line crosses the support of data density, which would cause a sharp change in the interpolated density. After decorrelation (right panel), we see that the support is roughly axis-parallel, and we expect that a simple linear interpolation should suffice.}
    \label{fig:scatter_plots}
\end{figure}

The primary motivation for density estimation methods is to address situations where the CWoLa assumption of statistical independence between the auxiliary features $\vec{x}$ and $m_{JJ}$ does not hold.
However, in the example considered above, $\vec{x}$ and $m_{JJ}$ were independent to a large degree.
In this section we explore how our strategies proposed above perform when $\vec{x}$ and $m_{JJ}$ are not independent.
Specifically, we artificially introduce dependence between $\vec{x}$ and $m_{JJ}$ via\footnote{We consider a non-polynomial dependence on $m_{JJ}$ instead of a linear one, as in \cite{Nachman:2020lpy, Hallin:2021wme}, because our decorrelation scheme below will be able to completely undo linear correlation, thus making the comparison not very useful.}
\begin{align}
    m_{J_1} &\to m_{J_1} + \log m_{JJ}, \\
    \Delta{m}_J &\to \Delta{m}_J + \log m_{JJ},
\end{align}
where all the masses are measured in units of TeV.

In this case, we immediately see a difficulty with our proposed interpolation method.
When $\vec{x}$ and $m_{JJ}$ are strongly dependent, the support of $p(m, \vec{x})$ can be of arbitrary shape in general, but the interpolation in \cref{eq:linterpol} implicitly assumes that for a fixed $\vec{x}$, $p(\vec{x}|m)$ does not vary too much as a function of $m_{JJ}$ across the SR.\footnote{Note that this is different from the CWoLa assumption since we only require weak dependence over the SR. In general this is easier to attain.}
This is illustrated in the left panel of \cref{fig:scatter_plots}, where a naive linear interpolation over the dashed line would result in an abrupt and unphysical drop in the interpolated density.
This situation can be handled automatically by NNs since they are able to perform more global interpolations, but  we need to be more careful when implementing the interpolation by hand.

It is clear from the above discussion that the quality of linear interpolation \cref{eq:linterpol} requires that $\vec{x}$ and $m_{JJ}$ be roughly independent \textit{over the SR}.
To achieve this, we perform the following simple ``decorrelation" procedure. For each feature $x$, consider the following transformation\footnote{We do not transform $m_{JJ}$ since this is a privileged  variable under the localized-signal assumption.}:
\begin{align}
    m &\to m \,, \\
    x &\to f(x, m) \,,
\end{align}
where $f$ is such that the transformation is bijective so that no information carried in $x$ is lost.
We can then search for $f$ such that the dependence between $x$ and $m_{JJ}$ within the SR, as measured by distance correlation, is minimized.
In particular, we consider $f$ belonging to a family of functions of the form
\begin{align}
    f(x, m) = a_0(m) + a_1(m) x.
\end{align}
Since $m_{JJ}$ lies within the SR which we assume to be small, we further parameterize the coefficients $a_0$ and $a_1$ as\footnote{Without loss of generality, we can take the constant term in $a_0$ to be zero and the constant term in $a_1$ to be 1, since distance correlation remains invariant under such a choice.}
\begin{align}
    a_0(m) = \alpha m + \beta m^2, \quad
    a_1(m) = 1 + \gamma m + \delta m^2 \,.
\end{align}
To summarize, we search for values of $\alpha, \beta, \gamma$ and $\delta$ that minimize the corresponding distance correlation.
This minimization is performed using the L-BFGS method~\cite{liu89} implemented in \texttt{SciPy}~\cite{2020SciPy-NMeth}.

The right panel of \cref{fig:scatter_plots} shows the scatter plot of $m_{J1}$ against $m_{JJ}$ over the SR after our decorrelation procedure.
Visually we can observe that the support of data density is now parallel to the $m_{JJ}$-axis, and numerically we can achieve a distance correlation of order \num{e-4} between auxiliary features and $m_{JJ}$ over the SR.
This signals the success of our decorrelation scheme.

\begin{figure}
    \centering
    \begin{subfigure}{0.49\textwidth}
        \centering
        \includegraphics[width=\textwidth]{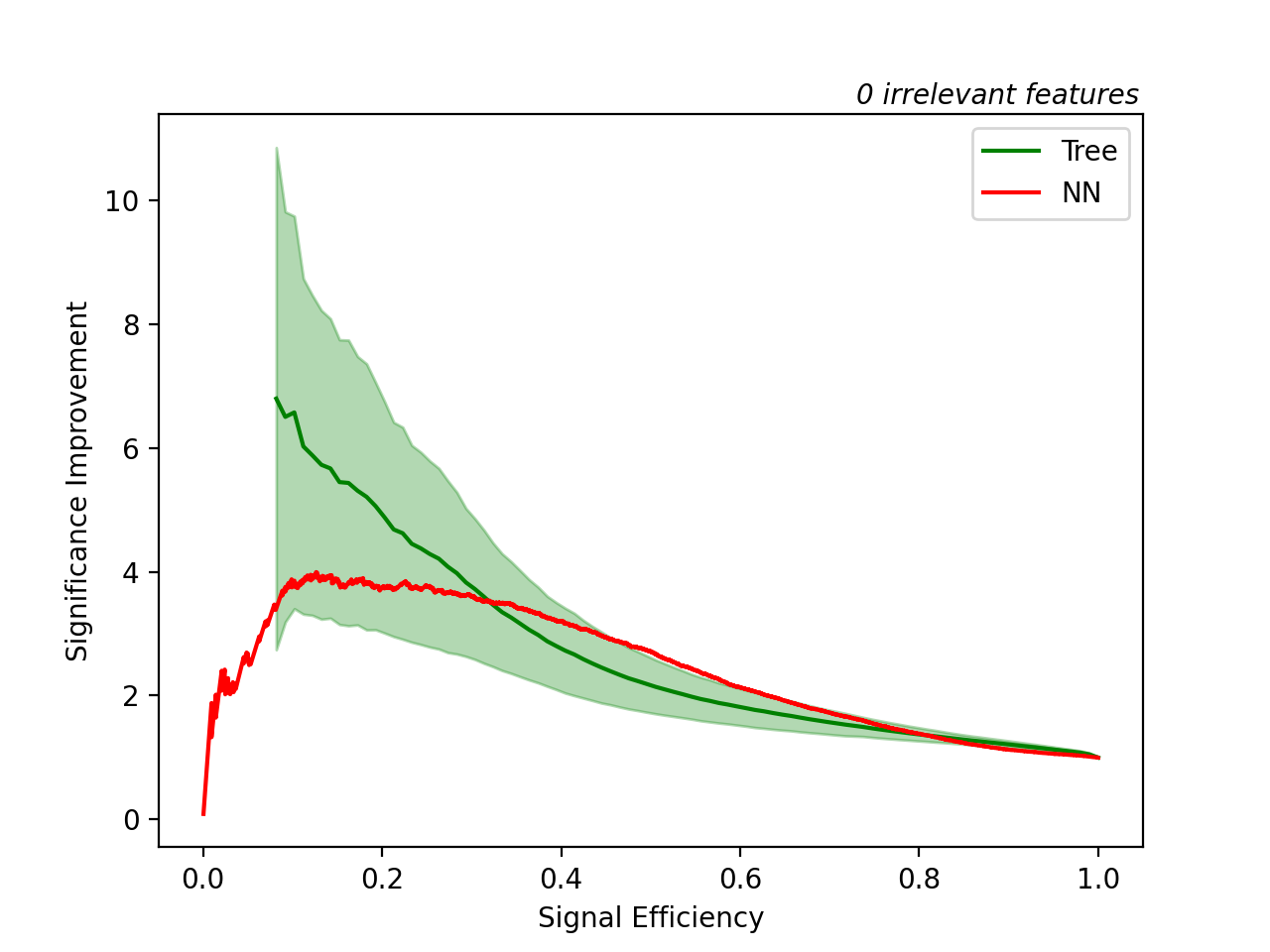}
    \end{subfigure}
    \hfill
    \begin{subfigure}{0.49\textwidth}
        \centering
        \includegraphics[width=\textwidth]{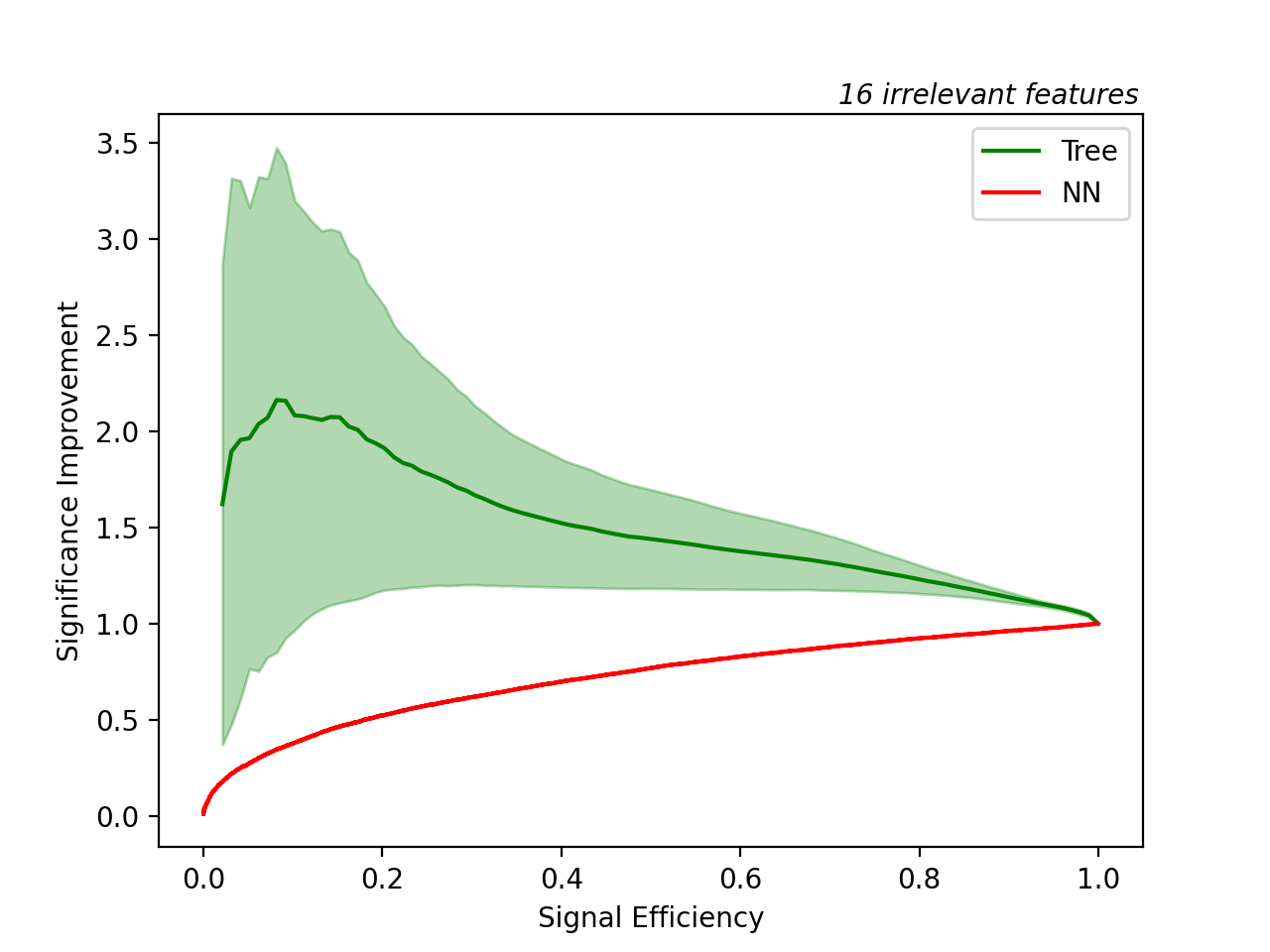}
    \end{subfigure}
    \caption{The SIC curves for NN-based (red) and tree-based (green) density-estimation algorithms applied to the log-shifted dataset. The cases with 0 (left panel) and 16 (right panel) irrelevant features are shown for comparison. The error bands are defined in the same way as in \cref{fig:SIC_ANODE}.}
    \label{fig:SIC_ANODE_shifted}
\end{figure}

With decorrelation carried out, the rest of the algorithm remains the same as in the previous section.
In \cref{fig:SIC_ANODE_shifted} we compare the SIC curves of the NN-based and tree-based algorithms applied to the log-shifted dataset.
We observe that the tree-level algorithm still greatly outperforms the NN-based method when irrelevant features are added.
At the same time, we also note that the performance of the tree-based algorithm is not as robust with respect to addition of irrelevant features as in the unshifted case (see \cref{sec:perf}).
This is likely due to the decorrelation procedure above, which by chance will find non-zero $(\alpha, \beta, \gamma, \delta)$ such that the in-sample distance correlation between $m_{JJ}$ and the transformed irrelevant feature is minimized.
This effect can in principle be mitigated by more rigorous cross-validation technique, but we do not pursue this point here.

While the simple approach to decorrelation and interpolation taken in this paper is effective, it may be seen as somewhat \emph{ad hoc}.
Many more elaborate interpolation methods exist in the literature (\emph{e.g.}, Gaussian process regression, high dimensional splines~\cite{hastie01statisticallearning}), which may further improve the performance and robustness of our algorithm.
We leave the exploration of such methods for future work.

\section{Discussion and Conclusions}
\label{sec:conc}

In this work, we have presented two tree-based approaches to detect anomalies in the presence of irrelevant features.
Anomaly detection methods are already starting to be used in LHC analyses, with searches based on CWoLa hunting at ATLAS already released~\cite{ATLAS:2020iwa}.
Since BDT-based methods are already used in experimental analyses, we hope that our methods would be readily able to be adopted and calibrated for experimental use.
We first considered a CWoLa-inspired method, and showed that boosted decision trees are more robust to irrelevant features compared to neural networks.
By exploiting the inherent feature selection of decision trees, the BDT-based classifier maintained good performance even with the addition of significantly more irrelevant than discriminating auxiliary features.

In analogy to density estimation methods like ANODE, we proposed using tree-based models paired with a copula transformation and interpolation step.
By estimating the marginal and copula densities separately, irrelevant features can be factorized out of the likelihood ratio assuming their mutual independence.
Even when this is not the case, we observe that the resulting reduction in significance improvement still leaves the tree-based approach much less sensitive to the presence of these features.
Our results demonstrated the promising performance of the tree-based density estimator compared to normalizing flows, especially in higher dimensionality with many irrelevant features.
The tree-based model allows for a simple and effective linear interpolation scheme for estimating the background density.

Recently, \cite{Finke:2023ltw} also explored the use of BDTs for anomaly detection in high-energy collider analyses.
This study includes a larger and more physical set of irrelevant features, while also finding increasingly improved performance and greater stability as irrelevant features are added during training.
However, it assumes that a perfect sample of the background signal is available and does not deal with the extrapolation of such a model into a resonant region, as we do.
We thus view our results as complementary and together making a compelling case for the application of tree-based methods to anomaly detection.

Overall, tree-based methods seem well-suited for anomaly detection tasks when operating on high-level observables with potential irrelevant features.
These naturally lend themselves to presentation as tabular data.
The techniques presented here could find useful application in collider searches and other physics analyses aiming to be robust against the embedding of low-dimensional signals in high-dimensional feature spaces.
More advanced interpolation schemes than what we consider here might improve the performance and stability of the density-based approach, while exploring other tree-based algorithms like Bayesian Additive Regression Trees might improve the overall fidelity of the learned functions.
We leave these possibilities to future work.

\section*{Acknowledgments}
We would like to thank Ben Nachman and David Shih for useful discussions.
This research is supported by the NSF grant PHY-2014071. 
YCS is partially supported by the Boochever Fellowship at Cornell University. 

\bibliographystyle{utphys}
\bibliography{ref}

\appendix
\section{Hyperparameter Tuning for \texttt{xgboost}}
\label{app:tune}

Here we provide details regarding hyperparameter tuning for the \texttt{xgboost} model used in CWoLa method.
The hyperparameters we choose to optimize are as follows\footnote{More details about these hyperparameters can be found in \cite{Chen_2016}}:
\begin{itemize}
    \item \texttt{n\_estimators}: this controls the number of boosting rounds
    \item \texttt{max\_depth}: this controls how complex the base tree learner is by limiting how deep each tree can be
    \item \texttt{eta}: this controls how much each tree contributes in building the ensemble
    \item \texttt{alpha}: $L_1$ regularizer on weights of the model
    \item \texttt{lambda}: $L_2$ regularizer on weights of the model
\end{itemize}
\texttt{xgboost} has a lot of other parameters, but here we choose to focus on these few because (i) \texttt{n\_estimators}, \texttt{max\_depth} and \texttt{eta} are known to have the most impact on the model's performance, and (ii) \texttt{alpha} and \texttt{lambda} explicitly control the model's weights, and therefore they have a direct impact on how much the model will overfit, which is exactly our concern here.
In addition, we (arbitrarily) fix the \texttt{subsample} parameter, which measures how much of the training data is used in fitting each individual tree, to be 0.75.
In principle one can also include it in the hyperparameter search but our empirical results show that the final performance is not very dependent on its exact value.
We use default values for all other hyperparameters.

To search for the optimal hyperparameters, we perform Bayesian optimization on the 10-fold cross-validation score, which we define to be the true positive rate of SR-SB labels at a fixed false positive rate of \num{e-3}.
Specifically, the Bayesian optimization is carried out using the \texttt{gp\_minimize} function in the \texttt{scikit-optimize} library, with default settings except we reduce the number of calls to 30 in order to save time.

\section{Hyperparameters of Boosted Density Estimation Tree Algorithm}
\label{app:BDTpar}

Here we describe briefly meanings of some of the hyperparameters used when training on the LHCO dataset.
Please refer to \cite{awaya2023unsupervised} for details.

\begin{itemize}
    \item \texttt{n\_estimators}: number of boosting rounds
    \item \texttt{max\_depth}: maximum depth each base tree learner can grow to
    \item \texttt{lr}: global shrinkage parameter that helps smooth out density learned during each boosting round. When it is equal to 0, each tree returns the uniform base distribution (no learning); when it is equal to 1, each tree returns the empirical distribution (most aggressive learning).
    \item \texttt{gamma}: amount of node-specific shrinkage. When it is 0, only the global learning rate \texttt{lr} is used; when it is a positive real number, the amount of shrinkage for each node grows as its volume in feature space decreases.
\end{itemize}

\section{Mutually Dependent Irrelevant Features}
\label{app:mutual}

\begin{figure}
    \centering
    \includegraphics[width=0.7\textwidth]{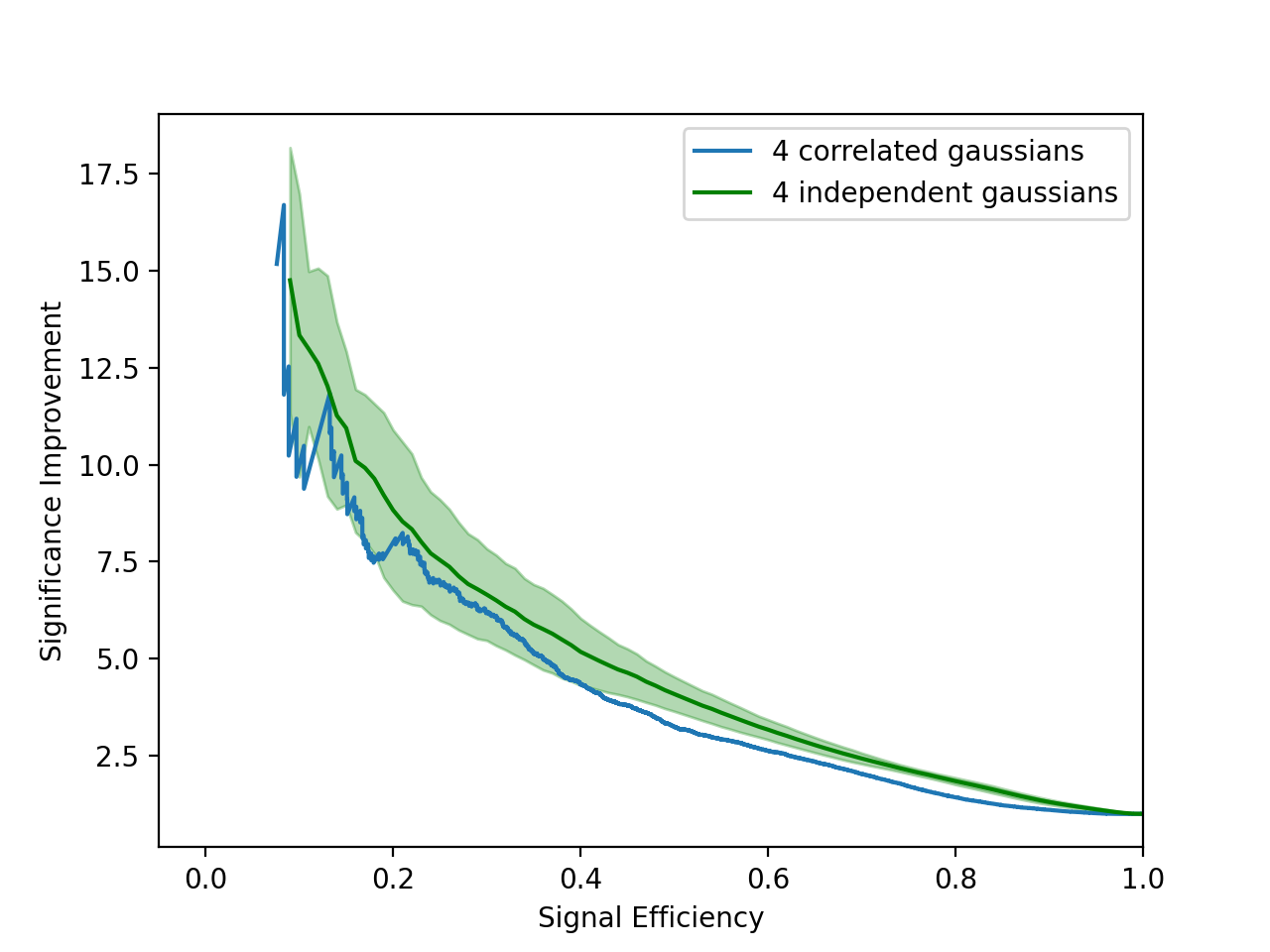}
    \caption{SIC curve of the tree-based density estimation algorithm with mutually correlated irrelevant features (blue), compared to the baseline case of mutually independent irrelevant features studied in \cref{sec:denest} (green).}
    \label{fig:anodemcorr}
\end{figure}

In the baseline model used throughout this paper, the irrelevant features enjoy the extra property that they are mutually independent, see \cref{eq:ifeatdist}.
While this extra property has no bearing on the CWoLa hunting method, it does affect our use of copula in \cref{sec:denest}: If the irrelevant features $\vec{y}$ are mutually independent, the copula density $c$ becomes independent of $\vec{y}$.
While such independence was not hardwired into our algorithm, it can potentially make the copula density easier to learn, and one might wonder how robust the algorithm is if irrelevant features are mutually dependent.

To test this, we rotate the original irrelevant features by a random matrix $A$:
\begin{align}
    \vec{y}^{~\prime} = A \vec{y} \,.
\end{align}
Here $A$ is constructed by independently sampling each of its elements from the standard normal distribution.
The elements of the rotated irrelevant feature vector $\vec{y}^{~\prime}$ are now mutually dependent.
We then apply the tree-based density estimation algorithm described in \cref{sec:denest} to the dataset $(m_{JJ}, \vec{x}, \vec{y}^{~\prime})$, where $\vec{x}$ are the relevant auxiliary features.

The resulting SIC curve is shown in \cref{fig:anodemcorr}.
We can see that our method's performance is very similar to the case considered in the main text, demonstrating that the method's performance is not reliant on the factorization property of \cref{eq:ifeatdist}.
In other words, the BDT is able to learn the non-trivial copula function involving irrelevant features well enough to not cause any degradation in the overall performance.
In future work, it would be interesting to further test this aspect of the algorithm in realistic physical applications of anomaly detection.  

\end{document}